\documentclass[article]{aastex631}

\definecolor{DarkerGreen}{rgb}{0.0,0.7,0.1}

\newcommand{\eisautofit}{{\tt eis\_auto\_fit.pro}}
\newcommand{\icsf}{{\tt icsf.pro}}

\newcommand{\solarsoft}{{\tt solarsoft}}

\usepackage{appendix,amsmath,ulem}
\received{}
\revised{}
\accepted{}

\shorttitle{Flows in active regions of TR}
\shortauthors{}
\begin{document}
\title{Center to limb variation of transition region Doppler shift in active regions}
\correspondingauthor{Abhishek Rajhans}
\email{abhishek@iucaa.in}
\author[0000-0001-5992-7060]{Abhishek Rajhans}
\author[0000-0003-1689-6254]{Durgesh Tripathi}
\affiliation{Inter-University Centre for Astronomy and Astrophysics, Post Bag - 4, Ganeshkhind, Pune 411007, India}
\author[0000-0002-3869-7996]{Vinay L. Kashyap}
\affiliation{Center for Astrophysics $|$ Harvard \& Smithsonian, 60 Garden St.\ Cambridge MA 02138, USA }

\author[0000-0003-2255-0305]{James A. Klimchuk}
\affiliation{NASA Goddard Space Flight Center, Solar Physics Laboratory, Code 671, Greenbelt, MD 20771, USA}
\author[0000-0002-7184-8004]{Avyarthana Ghosh}
\affiliation{Inter-University Centre for Astronomy and Astrophysics, Post Bag - 4, Ganeshkhind, Pune 411007, India}
\affiliation{Tata Research Development and Design Center, Tata Consultancy Services Ltd., India
}
\begin{abstract}

Studying Doppler shifts provides deeper insights into the flow of mass and energy in the solar atmosphere. We perform a comprehensive measurement of Doppler shifts in the transition region and its center-to-limb variation (CLV) in the strong field regions ($|\textbf{B}| \geq$ 50 G) of 50 active regions (ARs), using the \ion{Si}{4} 1394~{\AA} line recorded by the Interface Region Imaging Spectrometer(IRIS). To locate the ARs and identify strong field regions, we have used the magnetograms obtained by the Helioseismic and Magnetic Imager (HMI). We find that in strong field regions, on average, all the ARs show mean redshifts ranging between 4{--}11~ km/s, which varies with ARs. These flows show a mild CLV, with sizable magnitudes at the limb and substantial scatter at the mid-longitude range. Our observations do not support the idea that redshifts in the lower transition region (T $<\sim$ 0.1 MK) are produced by field-aligned downflows as a result of impulsive heating and warrant alternative interpretation, such as downflow of type-\rm{II} spicules in the presence of a chromospheric wall created by cooler type-\rm{I} spicules.

\end{abstract}

\section{Introduction}\label{sec:intro}
The heating of the solar atmosphere continues to be a challenging problem. Though magnetic fields are known to be responsible, the exact mechanism for energy dissipation and the transport of mass and energy across different layers of the atmosphere remains elusive. A possible explanation is the heating of the solar corona by impulsive events \citep[see for a review][]{klimchuk2006}. Impulsive heating results in the evaporation of chromospheric plasma along the loops into the corona, followed by draining and condensation. Hence, studying flows in different layers of the solar atmosphere sheds valuable insights into the heating and possible ways these layers may be coupled with each other.

Observations show that the transition region has a ubiquitous presence of redshifts (downflows). Early observations from Orbiting Solar Observatory \citep[OSO-8;][]{bruner1977}, the Naval Research Laboratory (NRL) normal incidence spectrograph onboard Skylab (S082-B), NRL High-Resolution Telescope and Spectrograph \citep[HRTS;][]{bartoejdf1975}, and the Ultra-Violet Spectrometer and Polarimeter  \citep[UVSP;][]{woodgate1980} onboard the Solar Maximum Mission \citep[SMM;][]{simnett1981} show downflows in the range of 5{--}20~km~s$^{-1}$ in ultraviolet spectral emission lines from bright regions in the chromosphere and the transition region \citep[also see][]{lemaire1978, bruckenerbartoe1980, gebbie1980, lites1980, bruckener1981, athaygurman1982,athaygurman1983, derekp1982, rottman1982, brekke1993, achour1995}. Moreover, transition region downflows in the range of 80{--}100~km~s$^{-1}$ have also been reported in small regions within active regions. However, due to their rare occurrence, these are considered to be associated with transients \citep{nicolas1982, dere1984}. 

Studies with similar scientific goals have also been performed using observations from Solar Ultraviolet Measurements of Emitted Radiation \citep[SUMER;][]{wilhelm1995}, the Coronal Diagnostic Spectrometer \citep[CDS;][]{harrison1995} onboard SOlar and Heliospheric Observatory \citep[SOHO;][]{domingo1995}, EUV Imaging Spectrometer \citep[EIS;][]{culhaneharra2007} onboard Hinode \citep{kosugi2007}, and Interface Region Imaging Spectrograph \citep[IRIS;][]{depointeau2014}. \cite{teriaca1999} used observations from SUMER to show downflows in the active regions to be ranging from $\sim$ 0 km s$^{-1}$ at log[T(K)] = 4.3 to $\sim$ 15 km s$^{-1}$ at log[T(K)] = 5.0. At log[T(K)] = 5.8 blueshifts $\sim$ 8 km s$^{-1}$ are observed. Further studies on plasma flows were conducted using observations from EIS \citep[see, e.g.,][]{delzanna2008,brookswarren2009,tripathi2009,tripathi2012,dadashi2011,gupta2015,ghosh2017} in warm loops as well as moss regions (transition region counterpart of hot loops). Persistent downflows were reported across the range of temperature EIS observed, $\log\,T=4.0$ to 5.0. However, lower transition region spectral lines like \ion{O}{4}, \ion{O}{5}, and \ion{Mg}{5} were very weak in the observations \citep{young2007}.

A possible explanation for these downflows can be impulsive heating occurring in the solar corona \citep[see, e.g.,][for a review]{klimchuk2006, reale2014, klimchuk2015}. In this scenario, the redshift is due to field-aligned downflows of cooling and draining plasma that were pushed up in the coronal loops due to chromospheric evaporation. If the flows have a random orientation relative to vertical, then they should show, on average, a center-to-limb variation (CLV) and vanish as one approaches the limb. However, \cite{feldman1982} found almost no CLV observed in data from NRL onboard Skylab(S082-B). They tracked two active regions as they traversed across the solar disk to study the Doppler shifts in the temperature range $\log\,T = 4.7${--}5.0, and found the downflows to be in the range of 4{--}17~km s$^{-1}$. Moreover, the redshifts extended out to the limb. \cite{klimchuk1987,klimchuk1989} used UVSP data and found similar results in measuring Doppler shifts relative to the average over the full raster. 

The observation of persistent downflows was explained by \cite{antiochos1984} as a signature of field-aligned flows due to condensation. Moreover, to explain the absence of CLV and non-diminishing flows at the limb, \cite{antiochos1984} introduced the idea of a \textit{chromospheric well}, which is formed due to the enhanced localized pressure due to impulsive heating. Under this scenario, the absence of CLV naturally arises due to projection effects. However, there is a drawback to this scenario. Under impulsive heating, field-aligned hydrodynamical simulations show downflows with much lower amplitude than those observed at similar temperatures. For example, the downflows in the \ion{Fe}{8} line formed at an approximate temperature of 0.4~MK is $\sim$0.9~km~s$^{-1}$ \citep[see, e.g.,][]{lopez2018, lopez2022}. Under the assumption of constancy of mass flux and pressure along a given flux tube in the transition region, the peak formation temperature of \ion{Si}{4} and \ion{Fe}{8} (forming at different heights in the transition region) imply that the speed of downflows in \ion{Si}{4} line should be less than $\lesssim$0.1~km~s$^{-1}$, which is about two orders of magnitude lower than observed velocities in the lower transition regions.

\cite{ghosh2019, GhoTK_2021} studied the Doppler shift and non-thermal velocities in \ion{Si}{4} line and their CLV for a single active region as it traversed the central meridian. They used the IRIS instrument, which provides regular spectroscopic observations of the transition region in \ion{Si}{4} line, with an accuracy of about 1~km s$^{-1}$. Moreover, the presence of multiple spectral lines due to neutral and single ionized ions provides the best opportunity to perform wavelength calibration and measure and characterize flows in the transition region. \cite{ghosh2019} found that the strong field regions of active regions (where magnetic field strengths are larger than 50 G) were redshifted by 5{--}10~km s$^{-1}$ and showed evidence of some CLV but less than expected for nearly vertical flows.

To mitigate the discrepancy between the observed redshifts and those obtained from hydrodynamical simulations, \cite{ghosh2019} suggested that the downflows observed in the transition regions are very likely related to the downflow of type-II spicules. They proposed the idea of a \textit{chromospheric wall} formed by cold spicules heated to a temperature of about 10$^{4}$ K in the vicinity of hot spicules, which get heated to 10$^{5}$ K. They argued that the optical depth of surrounding cold spicules is close to but less than unity, hence, allowing some center to limb variation in \ion{Si}{4} line. 

We note that \cite{ghosh2019} performed the Doppler measurements for a single active region while it crossed the central meridian. Additionally, the coverage of radius vectors (fractional distance to the limb from the disk center) was limited. Only eight values were covered in the range of -0.8{--}0.9 (0 denotes the disk center, and +1(-1) are the maximum values of the radius vector in the eastern (western) limb). The study assumed that AR evolution does not affect how flow velocities and directions might change. Hence, it is important to study a wide range of active regions to improve the longitude range and characterize the actual behavior of flow velocities. In this work, instead of tracking active regions, we  perform a snapshot study by observing Doppler shifts in different active regions at instants to check if the findings of \cite{ghosh2019} for one active region are valid for an ensemble of active regions. This provides a statistically larger sample and better longitudinal coverage. In \S\ref{sec:observations} we describe the data from different instruments used in this study. In \S\ref{sec:analysis} we describe the various procedures involved in analyzing data from different instruments, \textit{viz} (i) wavelength calibration, (ii) coalignment of data from AIA-1600, HMI, and IRIS, (iii) identification of strong-field regions within the active regions, and (iv) computation of Doppler shifts in these regions and associated radius vector. We discuss the results for all active regions and their CLV in \S\ref{sec:clvds}. We summarize in \S\ref{sec:disc}. 

\section{Observations and Data}\label{sec:observations} 
To study the Doppler shifts, we have used IRIS observations. IRIS provides spectra and images with spatial resolutions varying between 0.33{\arcsec} and 0.4{\arcsec} and a cadence of up to 20 s for spectra and 10 s for images. The field of view (FOV) can extend to 175{\arcsec}$\times$175{\arcsec}. The spectra obtained allow us to resolve velocities of 1~km~s$^{-1}$. 

IRIS records a pair of \ion{Si}{4} lines at 1393.78~{\AA} and 1402.77~{\AA}, with peak formation temperature $10^{4.9}$~K. Under the optically thin conditions, the line at 1393.78~{\AA} is a factor of two stronger than that at 1402.77~{\AA} (\citet{dere1996,landi2013}; see however, \citet{GonV_2018, TriNID_2020}). Hence, following \cite{ghosh2019}, we use the line at 1393.78~{\AA} for our study. 

We have also used observations from Atmospheric Imaging Assembly \citep[AIA;][]{lemen2012} in 1600~{\AA} filter for co-alignment purposes. We aim to study the Doppler shifts in the two major polarities of the ARs. Hence to identify the two polarities, we have used the line of sight (LOS) magnetograms obtained from Helioseismic and Magnetic Imager \citep[HMI;][]{schou2012a,schou2012b}.

\begin{figure}[ht]
\centering 
\includegraphics[width=0.8\linewidth]{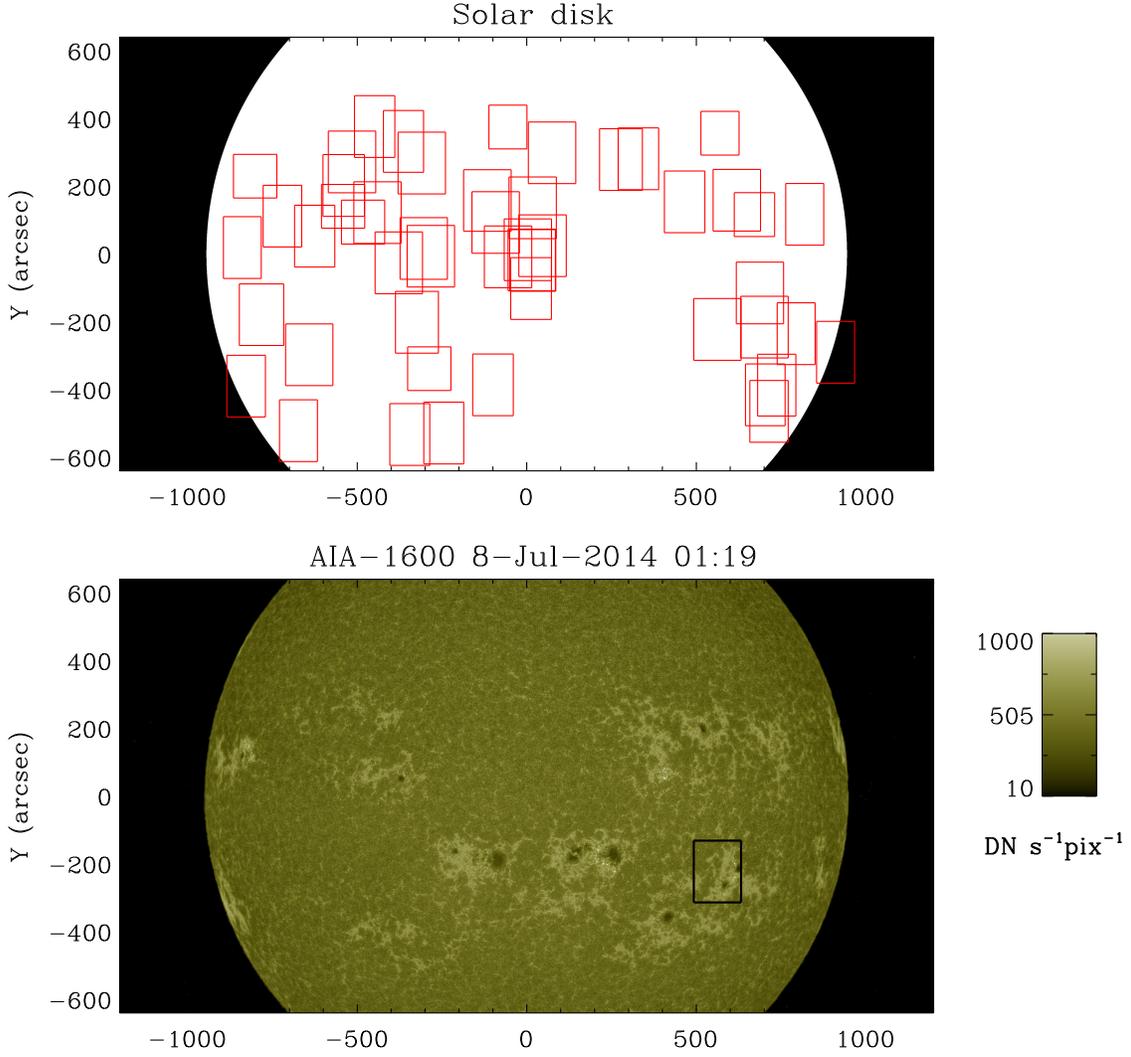} 
\caption{Top panel: Solar disk plotted with the red boxes showing the field of view (FOV) of all the IRIS rasters studied in the paper. Bottom panel: AIA-1600~{\AA} image taken on 8$^{th}$ of July 2014. The black box shows the FOV of the IRIS raster of the exemplar case discussed in detail.}\label{fig:aiairis}
\end{figure}

To study the CLV of the Doppler shift, we have selected 50 active regions, listed in Table~\ref{table:listofars}, observed at various locations covering the full range of longitudes. The upper panel of Figure~\ref{fig:aiairis} displays the location of all the active regions over the solar disk. The lower panel of Figure~\ref{fig:aiairis} shows AIA-1600~{\AA} image on $8^{th}$ of July 2014 with the black box showing the field of view of the IRIS raster corresponding to the exemplar active region (Case 39; the first row of  Table~\ref{table:listofars}) that is described in detail.

\section{Data Analysis and results}\label{sec:analysis}

To measure the absolute Doppler shift, we need to perform wavelength calibration. Also, since HMI and IRIS observe the Sun from two different vantage points, a proper coalignment needs to be ensured. For this purpose, we coalign IRIS observations with those obtained using AIA~1600{\AA} passband. Since AIA and HMI are both onboard the Solar Dynamics Observatory (SDO), the magnetograms can be readily coaligned with that of AIA. Once we obtain a calibrated Doppler map in \ion{Si}{4} and coaligned magnetograms, we identify the pixels associated with strong field areas of the active region and deduce the average Doppler shift. Here, we discuss the above-mentioned procedure in detail for an exemplar case of active region AR~12104. IRIS provided observations of this region from 23:35~UT on 7$^{th}$ of July 2014 to 03:05~UT on 8$^{th}$ of July 2014. The spatial extent of the corresponding IRIS raster extended from 490 to 630 arcseconds along the x{--}axis and {--}310 to {--}130{\arcsec} along the y{--}axis. The position of the raster for the exemplar case is shown with yellow box in Figure~\ref{fig:aiairis}.

\begin{figure}[ht]
\centering 
\includegraphics[width=0.8\linewidth]{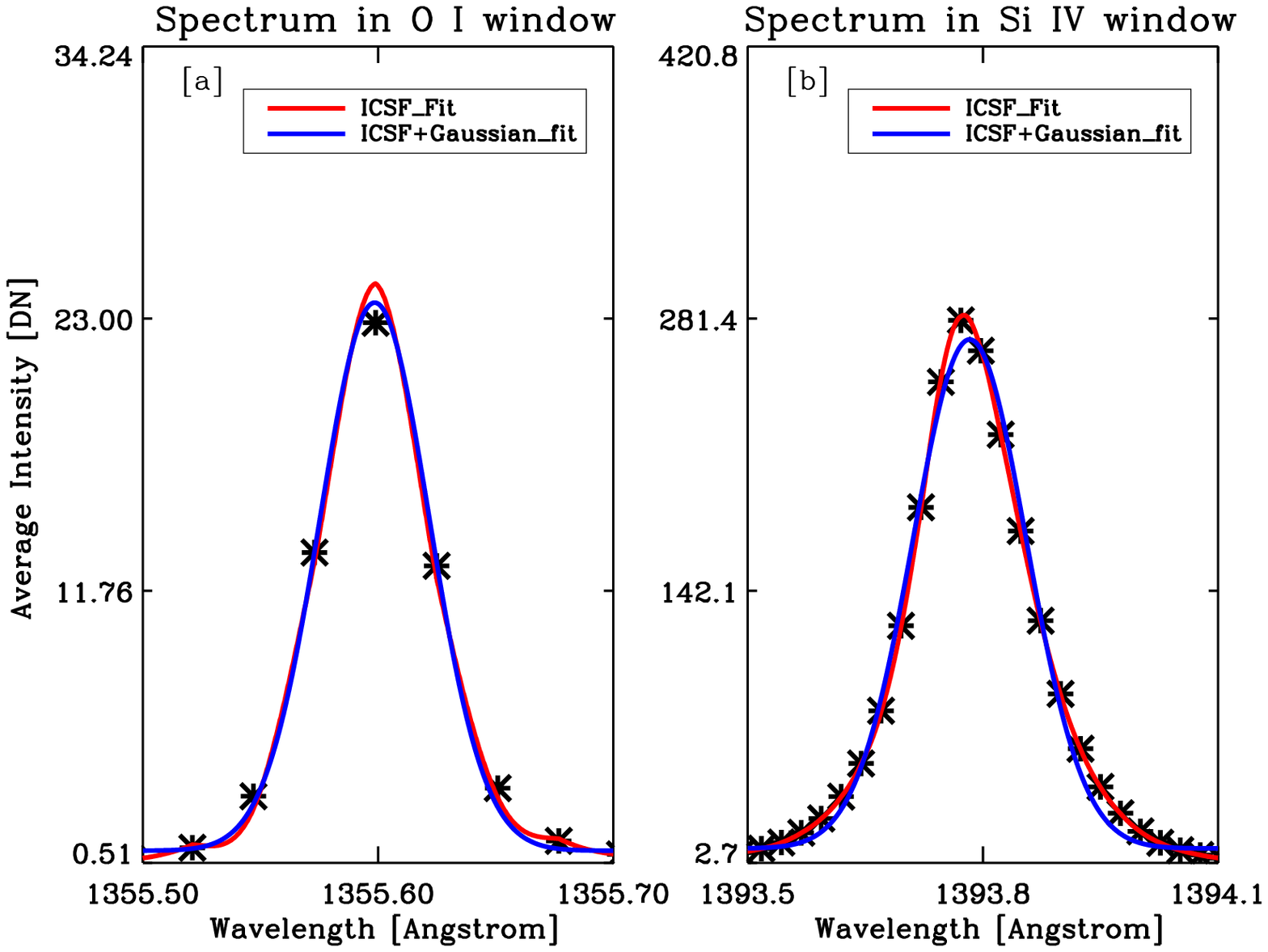} 
\caption{Spectrum obtained from IRIS in \ion{O}{1} (panel a) and \ion{Si}{4} (panel b) line windows. These spectra have been averaged over the full raster, for the exemplar case. The black asterisks show the data obtained from IRIS level2 fits files. Red curves denote the ICSF corrected spectrum. The blue curves show the Gaussian fit to the ICSF corrected spectrum. }\label{fig:o1si4spectrm}
\end{figure}

\subsection{Wavelength Calibration}\label{subsec:wvlcalib}

Wavelength calibration involves identifying average Doppler shifts in emission lines coming from neutral or singly ionized atoms, which are expected to be approximately at rest \citep{hassler1991}. Such neutral or singly ionized atoms are present in the photosphere or chromosphere. There are multiple lines such as \ion{Fe}{2}, \ion{O}{1}, and \ion{S}{1} present in IRIS spectral windows. Following \cite{ghosh2019}, we have used \ion{O}{1} (1355.6~{\AA}) line for performing wavelength calibration, which is a mid-chromospheric line in which velocities are small, and rarely exceeding 1.5 km/s for \ion{O}{1} line \citep{lin2015}. The average velocity can be assumed to be zero at the cost of finite random error. In this case, the spectrum should peak at the rest wavelength of the line. This should be the case in the ideal scenario because we expect atoms emitting these lines to be at rest. Any deviation in the peak of the spectrum from the rest wavelength should be due to instrumental effects, which need to be corrected.  

The average spectrum from a system of atoms at a finite non-zero temperature is Gaussian. However, directly fitting obtained spectrum with a Gaussian profile has limitations. The spectrum obtained by an instrument gives the average energy recorded in different wavelength bins, not the energy associated with the center of each bin. Even though, in the first approximation, the energy in the bin is associated with the central wavelength, it is valid only if the spectral line profile in the bin is linear. This certainly cannot be expected to be the case always. Consequently, to increase our accuracy in finding the line center, we have applied the method of Intensity Conserving Spline Fitting (ICSF) to the spectra using {\icsf} procedure \citep{klimchuk2016}. It preserves the total intensity in each spectral bin and performs a spline fitting to account for the line profile variation within the wavelength bin. Finally, the spectrum obtained after the application of ICSF procedure is fitted with a Gaussian using {\eisautofit} routine in {\solarsoft} \citep{freeland1998}. 

In Figure~\ref{fig:o1si4spectrm}, we plot the spectrum obtained in \ion{O}{1} (panel a) and \ion{Si}{4} (panel b) lines. These spectra have been averaged over the full raster. The black asterisks denote the original spectrum obtained from IRIS level2 fits files. The red curve represents the spectrum obtained after applying ICSF correction, and the blue curve is the final Gaussian fit to the ICSF correction. The  wavelength for \ion{Si}{4} is adjusted according to the difference between the laboratory rest wavelength of \ion{O}{1} and its observed wavelength of peak intensity in the raster-averaged spectrum. For the exemplar case, we find the wavelength at which the raster-averaged spectrum of \ion{O}{1} line peaks is 1355.5987~{\AA}, which is larger than the lab measurements of the rest wavelength, which is at 1355.5980 {\AA} as obtained from \cite{sandlin1986} and \cite{edlen1943}. The wavelength for \ion{Si}{4} is adjusted accordingly.

\subsection{Co-alignment of observations from IRIS, HMI, and AIA} \label{subsec:coalign}

\begin{figure}[ht]
\centering 
\includegraphics[width=0.765\linewidth]{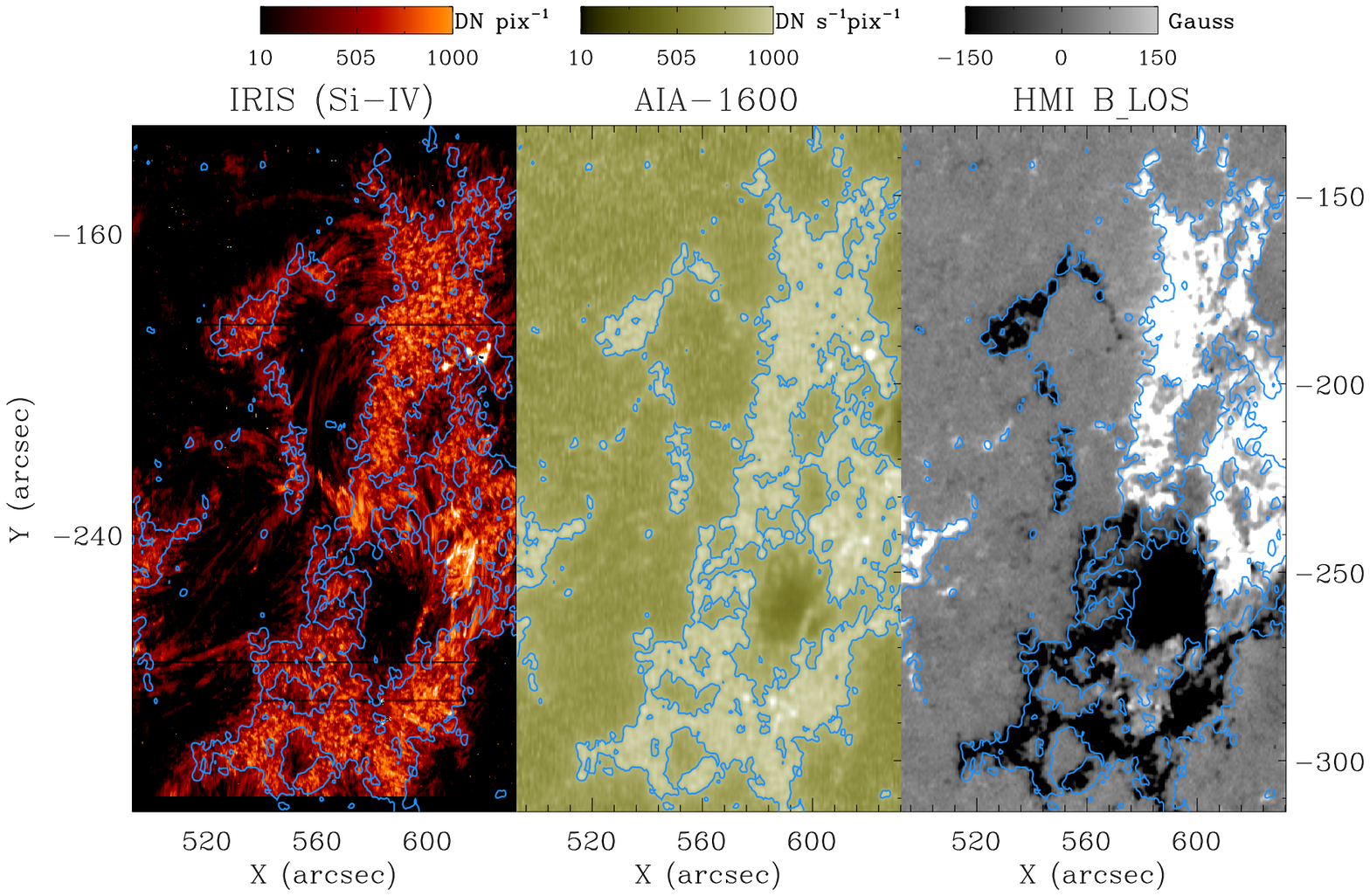} 
\caption{Intensity maps of emission in Si-IV line (left), artificial rasters of AIA-1600 (middle), and HMI LOS magnetogram (right). Contours of 250 DN s$^{-1}$ pix$^{-1}$ in AIA-1600 {\AA} filter are over-plotted.}\label{fig:aiahmiraster1} 
\end{figure}

\begin{figure}[hb]
\centering 
\includegraphics[width=0.765\textwidth]{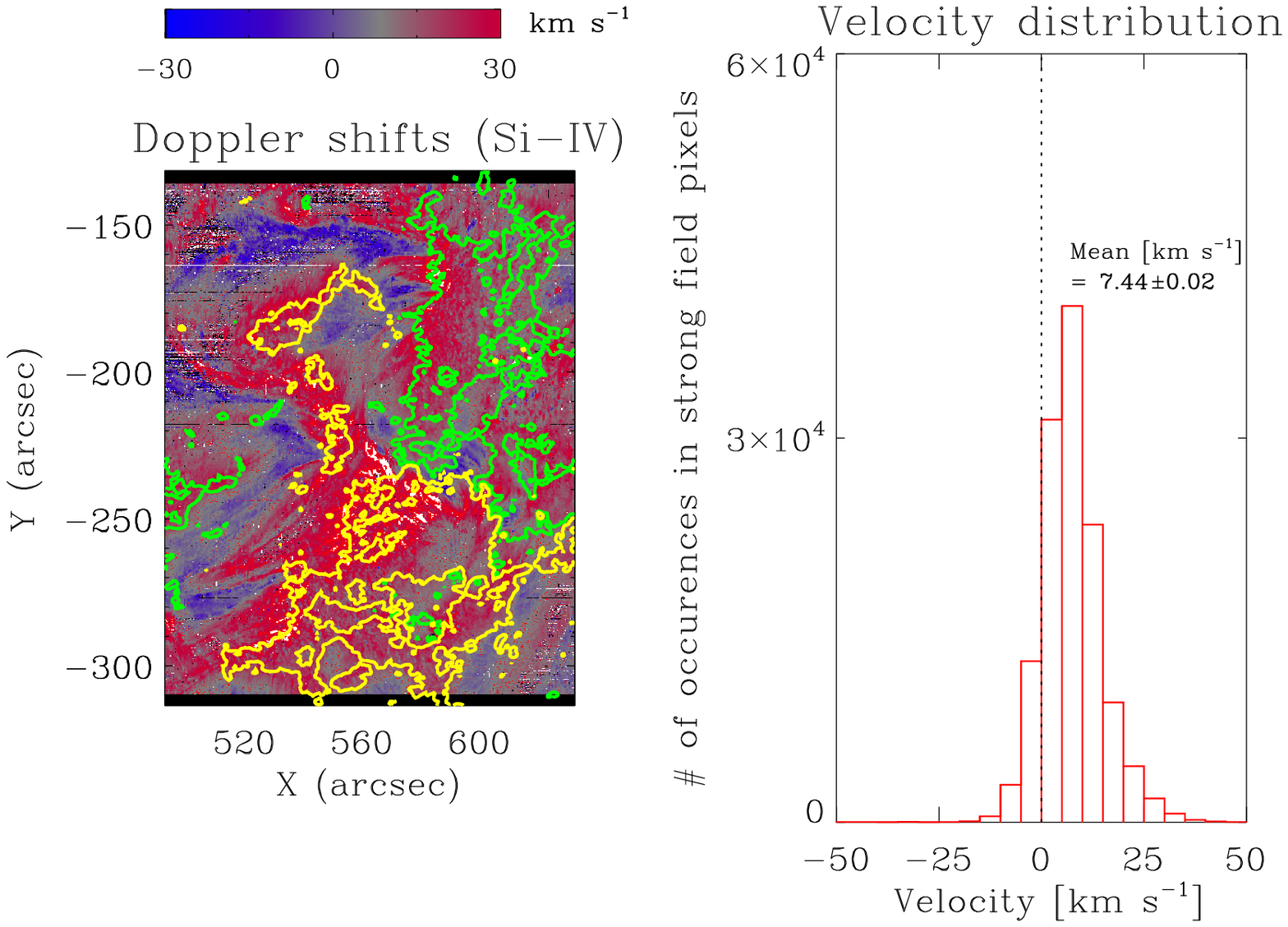} 
\caption{[Left] Velocity maps in \ion{Si}{4} line. The green and yellow contours in the right panel are of +50 and -50 Gauss, respectively.[Right] Histogram of velocities in strong field regions of active regions (where $|\textbf{B}| \geq$ 50 G). The dotted vertical line corresponds to 0 km s$^{-1}$. The average velocity in these pixels is $7.44 \pm 0.02$ km s$^{-1}$. } \label{fig:velocity} 
\end{figure}

For the purpose of coalignment of IRIS and HMI, we consider AIA observations taken at 1600~{\AA}, as this is closest in temperature to that is recorded by IRIS in \ion{Si}{4} line. We first make a data cube of AIA~1600~{\AA} images and LOS magnetograms of the region of interest during the entire duration of the raster. All AIA images and HMI LOS magnetograms have been coaligned with the IRIS slitjaw image in 1330~{\AA} taken at the closest time.  All the AIA images and LOS magnetograms in datacubes are then corrected for solar rotation with respect to the first AIA~1600~{\AA} image as the reference. We then create artificial AIA-1600 and HMI LOS magnetogram rasters corresponding to IRIS rasters to ensure proper coalignment.

Figure~\ref{fig:aiahmiraster1} (left panel) displays the intensity map obtained in \ion{Si}{4}. The middle and right panel displays the AIA~1600~{\AA} image and the LOS magnetogram obtained by artificial rastering. The over-plotted contours correspond to 250~DN~s$^{-1}$ pix$^{-1}$ in AIA~1600~{\AA} images. The excellent correspondence between the AIA contours on IRIS image and the magnetogram suggests a near-perfect coalignment of the data.

\subsection{Identification of active region and computing average Doppler shifts}\label{subsec:identifysfr}

After coaligning the data from different instruments and ensuring that we select the same structures from different data, we identify the strong field areas inside the active regions. \cite{klimchuk1987} identified the pixels in which the magnitude of the magnetic field exceeded 100 G. \cite{ghosh2019} on the other hand, used a magnitude of 50 G for the same purpose. \cite{ghosh2019} noted that the precise value is unimportant because the magnetic field strength decays rapidly outside the strong field regions. Consequently, the contours of magnetic fields of $\pm$ 100 G or $\pm$ 50 G are not very different. Following \cite{ghosh2019}, here we have used contours of $\pm$50~G to identify the strong field regions. 

We plot the velocity maps obtained in \ion{Si}{4} in the left panel of Figure~\ref{fig:velocity}. The over-plotted green and yellow contours correspond to $\pm$~50~Gauss, respectively, obtained from the magnetograms shown in Figure~\ref{fig:aiahmiraster1}.c. The right panel of Figure~\ref{fig:velocity} shows he histogram of velocity in such pixels. The average Doppler shifts in the strong field regions is 7.44 $\pm$ 0.02 km s$^{-1}$. We estimate the uncertainties by accounting for random errors due to variations in velocities and central wavelength of \ion{O}{1} across pixels identified as strong field regions. We have also taken into account the systematic errors due to an expected 3 mÅ [0.66 km/s] uncertainty in rest wavelength of \ion{O}{1} line used for wavelength calibration and a 0.1 pix dispersion uncertainty for \ion{Si}{4} [0.56 km/s]. For calculating errors, both random and systematic, their components have been added in quadrature. These procedures are discussed in detail in \cite{ghosh2019}. While the cumulative random error for this active region is $\pm$ 0.02 km s$^{-1}$, the total systematic error, which is assumed to be the same for all the regions, is $\approxeq$ 0.9~km~s$^{-1}$.

\subsection{Radius vector}\label{subsec:radiusvec}

We need to compute the radius vector of observed active regions to study the CLV of Doppler shifts. Radius vector is defined as the sine of the angle between the LOS and local vertical \citep{klimchuk1987}. A value of zero corresponds to disk center whereas positive(negative) radius vector represents longitudes to the east(west) of the central meridian.

We compute the radius vector of a given IRIS raster using its central pixel. If the central position of IRIS raster is [x,y] \arcsec, the radius vector is computed as $$\frac{\sqrt{x^{2}+y^{2}}}{R_{Sun}}$$ (where $R_{Sun}$ is 959 \arcsec). We multiply it by $\pm 1$ for the east(west) limb. For the exemplar case under consideration, the radius vector is 0.64.

\begin{table}[h!]
\centering
\caption{List of active regions studied. The file name of the IRIS rasters belonging to the different active regions studied is tabulated along with their radius vector (RV) and the average velocity $\pm$ in \ion{Si}{4} line ($V_{avg}$) in the pixels with strong magnetic fields($|\textbf{B}| \geq$ 50 G). The error cumulative random errors are also listed along with $V_{avg}$.}\label{table:listofars}
\begin{tabular}{ c c c c | c c c c}
\hline
\hline
Index & Fits file\footnote{The tabulated filename excludes the common part 'iris\_l2\_' on left and '\_raster\_t000\_r00000.fits' on right.} &  RV & V$_{avg}$  & Index & Fits file & RV & V$_{avg}$  \\ 
\hline
       0& 20201118\_153452\_3690108077
&    -0.94&      5.00$\pm$     0.10&      25&
20160227\_074513\_3620258078
&    -0.29&      8.37$\pm$    0.03\\
       1& 20200814\_054633\_3620108077
&    -0.89&      2.87$\pm$    0.07&      26&
20140701\_164900\_3820258196
&    -0.20&      9.73$\pm$    0.03\\
       2& 20180901\_175648\_3620108077
&    -0.88&      5.93$\pm$    0.07&      27&
20141001\_224938\_3800009396
&    -0.13&      10.3$\pm$    0.02\\
       3& 20141106\_024328\_3893010094
&    -0.86&      4.42$\pm$    0.06&      28&
20150223\_233348\_3800110096
&   -0.06&      9.75$\pm$    0.03\\
       4& 20140129\_200158\_3880010095
&    -0.82&      5.56$\pm$    0.05&      29&
20160228\_153411\_3620258078
&    0.01&      10.2$\pm$    0.03\\
       5& 20190605\_051739\_3620108077
&    -0.77&      7.14$\pm$    0.06&       30&
20150207\_041007\_3800256196
&    0.03&      9.95$\pm$    0.03\\
       6& 20140702\_003429\_3820259296
&    -0.74&      7.64$\pm$    0.02&      31&
20160330\_190439\_3600108078
&    0.05&      10.0$\pm$    0.03\\
       7& 20200105\_000818\_3690108077
&    -0.65&      7.41$\pm$    0.06&      32&
20171217\_031351\_3610108077
&     0.10&      8.34$\pm$     0.12\\
       8& 20180205\_113343\_3610108077
&    -0.64&      8.60$\pm$    0.09&      33&
20140701\_195503\_3820258196
&     0.14&      9.43$\pm$    0.03\\
       9& 20210923\_130908\_3620108077
&    -0.60&      7.69$\pm$    0.04&      34&
20150214\_150407\_3820256096
&     0.31&      9.48$\pm$    0.02\\
      10& 20160412\_020911\_3600108078
&    -0.60&      9.13$\pm$    0.02&      35&
20190309\_044823\_3620010077
&     0.40&      9.33$\pm$    0.05\\
      11& 20190507\_121550\_3620110077
&    -0.60&      6.06$\pm$    0.04&      36&
20150928\_170841\_3690092077
&     0.45&      9.92$\pm$    0.04\\
      12& 20191222\_045119\_3690108077
&    -0.60&      8.61$\pm$     0.17&      37&
20180601\_030145\_3620108077
&     0.45&      6.87$\pm$    0.03\\
      13& 20140916\_044847\_3893010094
&    -0.58&      8.17$\pm$    0.05&      38&
20190322\_105259\_3620108077
&     0.50&      9.42$\pm$    0.07\\
      14& 20171210\_021825\_3630108077
&    -0.56&      7.37$\pm$    0.03&      39&
20140707\_233530\_3824263396
&     0.64&      7.43$\pm$    0.02\\
      15& 20200801\_143730\_3620108077
&    -0.52&      5.25$\pm$    0.04&      40&
20160310\_163211\_3620258078
&     0.66&      9.68$\pm$    0.03\\
      16& 20141130\_070200\_3893010094
&    -0.50&      6.90$\pm$    0.06&      41&
20170306\_072447\_3620106076
&     0.69&      8.75$\pm$    0.05\\
      17& 20140815\_070803\_3800258196
&    -0.48&      9.51$\pm$    0.02&      42&
20160115\_120419\_3630008076
&     0.69&      6.49$\pm$    0.04\\
      18& 20141003\_044846\_3893260094
&    -0.46&      6.69$\pm$    0.06&      43&
20140215\_163205\_3800258296
&     0.71&      5.35$\pm$    0.02\\
      19& 20160413\_014409\_3600108078
&    -0.42&      9.39$\pm$    0.02&      44&
20140708\_192613\_3824263396
&     0.77&      7.04$\pm$    0.03\\
      20& 20201206\_090413\_3610108077
&    -0.40&      6.29$\pm$    0.08&      45&
20191122\_101544\_3690108077
&     0.84&      6.71$\pm$     0.21\\
      21& 20171025\_161338\_3630110077
&    -0.39&      6.64$\pm$    0.06&      46&
20200501\_145553\_3620108077
&     0.87&      9.19$\pm$     0.28\\
      22& 20170303\_021419\_3620106076
&    -0.39&      8.43$\pm$    0.04&      47&
20200801\_011722\_3620108077
&     0.88&      7.00$\pm$     0.10\\
      23& 20150222\_154645\_3800110096
&    -0.39&      9.07$\pm$    0.03&      48&
20200315\_235329\_3620108077
&     0.88&      6.75$\pm$     0.10\\
      24& 20160329\_061311\_3600108078
&    -0.31&      9.06$\pm$    0.03&      49&
20201205\_164951\_3610108077
&     0.98&      3.87$\pm$    0.07\\
\hline 
\end{tabular}
\end{table}

\begin{figure}[h]
\centering 
\includegraphics[width=0.8\linewidth]{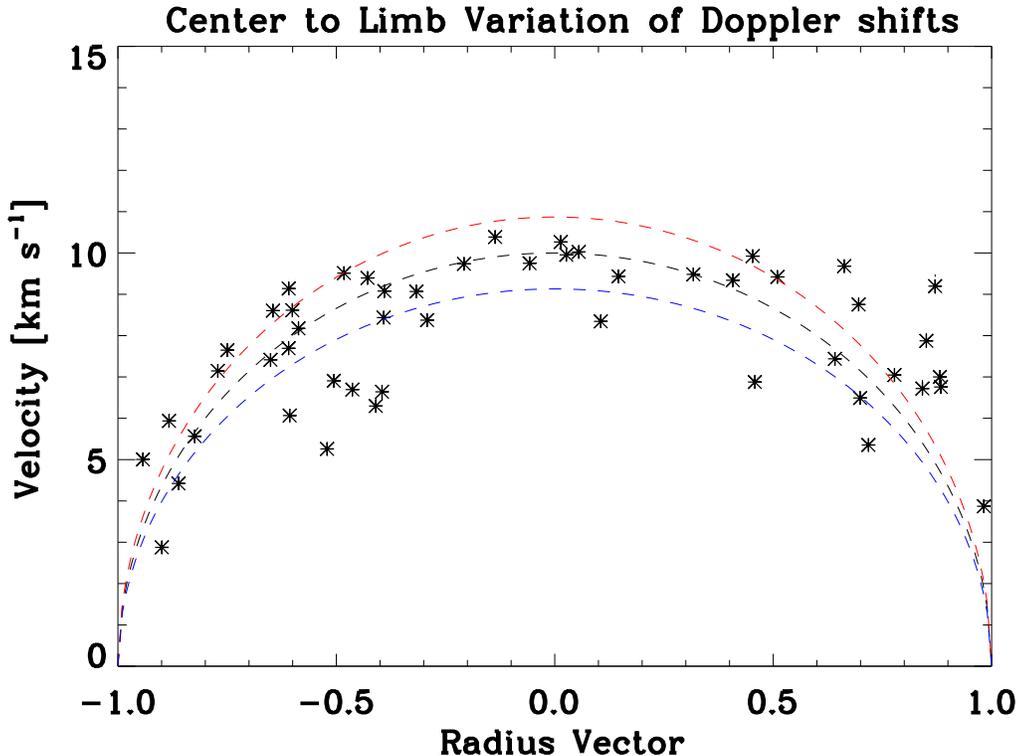} 
\caption{Measured Doppler shifts in the strong field regions of the active region as a function of radius vector shown with black asterisks. The dashed black curve shows the variation of Doppler velocity expected from the hypothetical vertical flow of $v_{vertical} =$ 10 km s$^{-1}$. The over-plotted blue and red dashed lines show the CLV of hypothetical vertical flows with velocities $v_{vertical}-\delta v_{sys}$ (blue), and $v_{vertical}+\delta v_{sys}$ (red). The random errors range from 0.01 km $s^{-1}$ to 0.2 km $s^{-1}$. Consequently, these errors are hardly visible.} \label{fig:clv} 
\end{figure}

\section{Center to Limb variation of Doppler shifts}\label{sec:clvds}

We carry out exactly the same analysis discussed above for 50 active regions listed in Table~\ref{table:listofars}. These active regions have been randomly selected so as cover radius vectors over the whole disk. The name of the analyzed iris level2 fits files, their radius vector (RV), the mean velocity (V$_{avg}$ with random error) in its strong field regions are given in Table~\ref{table:listofars}. These active regions have been arranged in ascending order of radius vectors. Case 39 corresponds to the exemplar case discussed in the previous section. Figure~\ref{fig:clv} plots the Doppler shifts as a function of radius vector. The black asterisks show the average Doppler shift in the strong field regions. 

We demonstrate the expected behavior of the line-of-sight Doppler shifts,
$$v_{LOS} = v_{0}\sqrt{\left[1-\left(\frac{r}{R}\right)^{2}\right]},$$
where $v_{0}$ is the mean velocity corresponding to an active region at disk center, and RV$=\frac{r}{R}$ is the radius vector.  The black dashed lines in Figure~\ref{fig:clv} show the trend for $v_{0}=10$~km~s$^{-1}$, which matches the measured average Doppler shift in the strong field region of the active region closest to disk center (at RV$=0.01$). The effect of systematic shifts from variations of $\pm$0.9~km~s$^{-1}$ are shown as red and blue dashed curves.  These curves show what kind of variation is expected {\sl if} $v_{0}$ were the same for all active regions.  In reality, the average flow velocity will vary across different active regions.  However, note that the scatter in the data points is similar to the expected systematic variations, suggesting that the variation in $v_{0}$ across active regions is not much larger than $\sim$1~km~s$^{-1}$.  Figure~\ref{fig:clv} is thus an illustration of how the data differs from the expected variation.  Some center-to-limb variation is clearly visible, consistent with the observations of \cite{ghosh2019} based on following a single active region over time.  Large departures from the expected trend are also seen as the RV approaches the limb, suggesting that an additional effect is responsible. Detailed modeling of these effects is beyond the scope of this work.

\section{Summary and Discussion}\label{sec:disc}
Here, we report on the most comprehensive measurements and analysis of AR Doppler shifts and CLV to date using IRIS spectral measurements of \ion{Si}{4}. For the purpose of co-alignment and to identify the strong field regions in active regions, we have used the observations from AIA and HMI, both onboard SDO. 

Similar to the results obtained by \cite{feldman1976}, \cite{klimchuk1987,klimchuk1989}, and \cite{ghosh2019}, we find that in lower transition region emissions, active regions are predominantly red shifted with velocities ranging between 4{--}11~km~s$^{-1}$. Moreover, the Doppler shifts show CLV, as was also reported by \cite{ghosh2019}. Note that the results obtained by \cite{ghosh2019} was based on the tracking of a single active region AR 12641 as it crossed from the center towards the limb. Here, we have studied 50 active regions located at different locations across the solar disk.

\citet{ghosh2019} proposed that the lower transition region redshifts are not  due to the draining of cooling coronal material but rather the main bodies of falling type~\rm{II} spicules. If these spicules have a random orientation relative to vertical, then their average redshift should exhibit the CLV expected of a vertical flow. To explain the weaker variation that is observed, Ghosh et al. proposed that absorption from interlaced cold type~\rm{I} spicules gives preferential weighting to type~\rm{II} spicules that are more closely aligned with the line of sight. These "selected" spicules have similar Doppler shift everywhere across the solar disk. A modest CLV occurs because the type~\rm{I} spicules are not totally opaque. \citet{ghosh2019} described the type~\rm{I} spicules as providing a chromospheric wall, which is slightly different from the chromospheric well proposed by \cite{antiochos1984}.

We thank the referee for careful reading and constructive comments. This research is partly supported by the Max-Planck Partner Group on the Coupling and Dynamics of the Solar Atmosphere of MPS at IUCAA. AR acknowledges financial support from University Grants Commission in form of SRF. VLK acknowledges support from NASA Contract NAS8-03060 to the Chandra X-ray Center, and the hospitality of IUCAA during several visits. The work of JAK was supported by the Internal Scientist Funding Model (competitive work package program) at Goddard Space Flight Center. AIA and HMI data are onboard SDO (NASA mission). IRIS is a small explorer mission of NASA, developed and operated by LMSAL.
\bibliography{clv}

\begin{thebibliography}{}
\expandafter\ifx\csname natexlab\endcsname\relax\def\natexlab#1{#1}\fi
\providecommand{\url}[1]{\href{#1}{#1}}
\providecommand{\dodoi}[1]{doi:~\href{http://doi.org/#1}{\nolinkurl{#1}}}
\providecommand{\doeprint}[1]{\href{http://ascl.net/#1}{\nolinkurl{http://ascl.net/#1}}}
\providecommand{\doarXiv}[1]{\href{https://arxiv.org/abs/#1}{\nolinkurl{https://arxiv.org/abs/#1}}}

\bibitem[{{Achour} {et~al.}(1995){Achour}, {Brekke}, {Kjeldseth-Moe}, \&
  {Maltby}}]{achour1995}
{Achour}, H., {Brekke}, P., {Kjeldseth-Moe}, O., \& {Maltby}, P. 1995, \apj,
  453, 945, \dodoi{10.1086/176454}

\bibitem[{{Antiochos}(1984)}]{antiochos1984}
{Antiochos}, S.~K. 1984, \apj, 280, 416, \dodoi{10.1086/162007}

\bibitem[{{Athay} {et~al.}(1983){Athay}, {Gurman}, \&
  {Henze}}]{athaygurman1983}
{Athay}, R.~G., {Gurman}, J.~B., \& {Henze}, W. 1983, \apj, 269, 706,
  \dodoi{10.1086/161080}

\bibitem[{{Athay} {et~al.}(1982){Athay}, {Gurman}, {Shine}, \&
  {Henze}}]{athaygurman1982}
{Athay}, R.~G., {Gurman}, J.~B., {Shine}, R.~A., \& {Henze}, W. 1982, \apj,
  261, 684, \dodoi{10.1086/160379}

\bibitem[{{Bartoe} \& {Brueckner}(1975)}]{bartoejdf1975}
{Bartoe}, J.~D.~F., \& {Brueckner}, G.~E. 1975, in Bulletin of the American
  Astronomical Society, Vol.~7, 432

\bibitem[{{Brekke}(1993)}]{brekke1993}
{Brekke}, P. 1993, \apjs, 87, 443, \dodoi{10.1086/191810}

\bibitem[{{Brooks} \& {Warren}(2009)}]{brookswarren2009}
{Brooks}, D.~H., \& {Warren}, H.~P. 2009, \apjl, 703, L10,
  \dodoi{10.1088/0004-637X/703/1/L10}

\bibitem[{{Brueckner}(1981)}]{bruckener1981}
{Brueckner}, G.~E. 1981, in Solar Active Regions: A monograph from Skylab Solar
  Workshop III, ed. F.~Q. {Orrall}, 113--127

\bibitem[{{Brueckner} {et~al.}(1980){Brueckner}, {Bartoe}, \&
  {Dykton}}]{bruckenerbartoe1980}
{Brueckner}, G.~E., {Bartoe}, J. D.~F., \& {Dykton}, M. 1980, in Bulletin of
  the American Astronomical Society, Vol.~12, 907

\bibitem[{{Bruner}(1977)}]{bruner1977}
{Bruner}, E.~C., J. 1977, Space Science Instrumentation, 3, 369

\bibitem[{{Culhane} {et~al.}(2007){Culhane}, {Harra}, {James}, {Al-Janabi},
  {Bradley}, {Chaudry}, {Rees}, {Tandy}, {Thomas}, {Whillock}, {Winter},
  {Doschek}, {Korendyke}, {Brown}, {Myers}, {Mariska}, {Seely}, {Lang}, {Kent},
  {Shaughnessy}, {Young}, {Simnett}, {Castelli}, {Mahmoud}, {Mapson-Menard},
  {Probyn}, {Thomas}, {Davila}, {Dere}, {Windt}, {Shea}, {Hagood}, {Moye},
  {Hara}, {Watanabe}, {Matsuzaki}, {Kosugi}, {Hansteen}, \&
  {Wikstol}}]{culhaneharra2007}
{Culhane}, J.~L., {Harra}, L.~K., {James}, A.~M., {et~al.} 2007, \solphys, 243,
  19, \dodoi{10.1007/s01007-007-0293-1}

\bibitem[{{Dadashi} {et~al.}(2011){Dadashi}, {Teriaca}, \&
  {Solanki}}]{dadashi2011}
{Dadashi}, N., {Teriaca}, L., \& {Solanki}, S.~K. 2011, \aap, 534, A90,
  \dodoi{10.1051/0004-6361/201117234}

\bibitem[{{De Pontieu} {et~al.}(2014){De Pontieu}, {Title}, \&
  {Carlsson}}]{depointeau2014}
{De Pontieu}, B., {Title}, A., \& {Carlsson}, M. 2014, Science, 346, 315,
  \dodoi{10.1126/science.346.6207.315}

\bibitem[{{Del Zanna}(2008)}]{delzanna2008}
{Del Zanna}, G. 2008, \aap, 481, L49, \dodoi{10.1051/0004-6361:20079087}

\bibitem[{{Dere}(1982)}]{derekp1982}
{Dere}, K.~P. 1982, \solphys, 77, 77, \dodoi{10.1007/BF00156097}

\bibitem[{{Dere} {et~al.}(1984){Dere}, {Bartoe}, \& {Brueckner}}]{dere1984}
{Dere}, K.~P., {Bartoe}, J. D.~F., \& {Brueckner}, G.~E. 1984, \apj, 281, 870,
  \dodoi{10.1086/162167}

\bibitem[{{Dere} {et~al.}(1996){Dere}, {Monsignori-Fossi}, {Landi}, {Mason}, \&
  {Young}}]{dere1996}
{Dere}, K.~P., {Monsignori-Fossi}, B.~C., {Landi}, E., {Mason}, H.~E., \&
  {Young}, P.~R. 1996, in American Astronomical Society Meeting Abstracts, Vol.
  188, American Astronomical Society Meeting Abstracts \#188, 85.01

\bibitem[{{Domingo} {et~al.}(1995){Domingo}, {Fleck}, \&
  {Poland}}]{domingo1995}
{Domingo}, V., {Fleck}, B., \& {Poland}, A.~I. 1995, \solphys, 162, 1,
  \dodoi{10.1007/BF00733425}

\bibitem[{{Edl{\'e}n}(1943)}]{edlen1943}
{Edl{\'e}n}, B. 1943, \zap, 22, 30

\bibitem[{{Feldman} {et~al.}(1982){Feldman}, {Doschek}, \&
  {Cohen}}]{feldman1982}
{Feldman}, U., {Doschek}, G.~A., \& {Cohen}, L. 1982, \apj, 255, 325,
  \dodoi{10.1086/159833}

\bibitem[{{Feldman} {et~al.}(1976){Feldman}, {Doschek}, \&
  {Patterson}}]{feldman1976}
{Feldman}, U., {Doschek}, G.~A., \& {Patterson}, N.~P. 1976, \apj, 209, 270,
  \dodoi{10.1086/154718}

\bibitem[{{Freeland} \& {Handy}(1998)}]{freeland1998}
{Freeland}, S.~L., \& {Handy}, B.~N. 1998, \solphys, 182, 497,
  \dodoi{10.1023/A:1005038224881}

\bibitem[{{Gebbie} {et~al.}(1980){Gebbie}, {Hill}, {Toomre}, {November},
  {Simon}, {Athay}, {Bruner}, {Rehse}, {Gurman}, {Shine}, {Woodgate}, \&
  {Tandberg-Hanssen}}]{gebbie1980}
{Gebbie}, K.~B., {Hill}, F., {Toomre}, J., {et~al.} 1980, in Bulletin of the
  American Astronomical Society, Vol.~12, 907

\bibitem[{{Ghosh} {et~al.}(2019){Ghosh}, {Klimchuk}, \& {Tripathi}}]{ghosh2019}
{Ghosh}, A., {Klimchuk}, J.~A., \& {Tripathi}, D. 2019, \apj, 886, 46,
  \dodoi{10.3847/1538-4357/ab43c4}

\bibitem[{{Ghosh} {et~al.}(2017){Ghosh}, {Tripathi}, {Gupta}, {Polito},
  {Mason}, \& {Solanki}}]{ghosh2017}
{Ghosh}, A., {Tripathi}, D., {Gupta}, G.~R., {et~al.} 2017, \apj, 835, 244,
  \dodoi{10.3847/1538-4357/835/2/244}

\bibitem[{{Ghosh} {et~al.}(2021){Ghosh}, {Tripathi}, \&
  {Klimchuk}}]{GhoTK_2021}
{Ghosh}, A., {Tripathi}, D., \& {Klimchuk}, J.~A. 2021, \apj, 913, 151,
  \dodoi{10.3847/1538-4357/abf244}

\bibitem[{{Gontikakis} \& {Vial}(2018)}]{GonV_2018}
{Gontikakis}, C., \& {Vial}, J.~C. 2018, \aap, 619, A64,
  \dodoi{10.1051/0004-6361/201732563}

\bibitem[{{Gupta} {et~al.}(2015){Gupta}, {Tripathi}, \& {Mason}}]{gupta2015}
{Gupta}, G.~R., {Tripathi}, D., \& {Mason}, H.~E. 2015, \apj, 800, 140,
  \dodoi{10.1088/0004-637X/800/2/140}

\bibitem[{{Harrison} {et~al.}(1995){Harrison}, {Sawyer}, {Carter}, {Cruise},
  {Cutler}, {Fludra}, {Hayes}, {Kent}, {Lang}, {Parker}, {Payne}, {Pike},
  {Peskett}, {Richards}, {Gulhane}, {Norman}, {Breeveld}, {Breeveld}, {Al
  Janabi}, {McCalden}, {Parkinson}, {Self}, {Thomas}, {Poland}, {Thomas},
  {Thompson}, {Kjeldseth-Moe}, {Brekke}, {Karud}, {Maltby}, {Aschenbach},
  {Br{\"a}uninger}, {K{\"u}hne}, {Hollandt}, {Siegmund}, {Huber}, {Gabriel},
  {Mason}, \& {Bromage}}]{harrison1995}
{Harrison}, R.~A., {Sawyer}, E.~C., {Carter}, M.~K., {et~al.} 1995, \solphys,
  162, 233, \dodoi{10.1007/BF00733431}

\bibitem[{{Hassler} {et~al.}(1991){Hassler}, {Rottman}, \&
  {Orrall}}]{hassler1991}
{Hassler}, D.~M., {Rottman}, G.~J., \& {Orrall}, F.~Q. 1991, Advances in Space
  Research, 11, 141, \dodoi{10.1016/0273-1177(91)90102-P}

\bibitem[{{Klimchuk}(1987)}]{klimchuk1987}
{Klimchuk}, J.~A. 1987, \apj, 323, 368, \dodoi{10.1086/165834}

\bibitem[{{Klimchuk}(1989)}]{klimchuk1989}
---. 1989, \solphys, 119, 19, \dodoi{10.1007/BF00146209}

\bibitem[{{Klimchuk}(2006)}]{klimchuk2006}
---. 2006, \solphys, 234, 41, \dodoi{10.1007/s11207-006-0055-z}

\bibitem[{{Klimchuk}(2015)}]{klimchuk2015}
---. 2015, Philosophical Transactions of the Royal Society of London Series A,
  373, 20140256, \dodoi{10.1098/rsta.2014.0256}

\bibitem[{{Klimchuk} {et~al.}(2016){Klimchuk}, {Patsourakos}, \&
  {Tripathi}}]{klimchuk2016}
{Klimchuk}, J.~A., {Patsourakos}, S., \& {Tripathi}, D. 2016, \solphys, 291,
  55, \dodoi{10.1007/s11207-015-0827-4}

\bibitem[{{Kosugi} {et~al.}(2007){Kosugi}, {Matsuzaki}, {Sakao}, {Shimizu},
  {Sone}, {Tachikawa}, {Hashimoto}, {Minesugi}, {Ohnishi}, {Yamada}, {Tsuneta},
  {Hara}, {Ichimoto}, {Suematsu}, {Shimojo}, {Watanabe}, {Shimada}, {Davis},
  {Hill}, {Owens}, {Title}, {Culhane}, {Harra}, {Doschek}, \&
  {Golub}}]{kosugi2007}
{Kosugi}, T., {Matsuzaki}, K., {Sakao}, T., {et~al.} 2007, \solphys, 243, 3,
  \dodoi{10.1007/s11207-007-9014-6}

\bibitem[{{Landi} {et~al.}(2013){Landi}, {Young}, {Dere}, {Del Zanna}, \&
  {Mason}}]{landi2013}
{Landi}, E., {Young}, P.~R., {Dere}, K.~P., {Del Zanna}, G., \& {Mason}, H.~E.
  2013, \apj, 763, 86, \dodoi{10.1088/0004-637X/763/2/86}

\bibitem[{{Lemaire} {et~al.}(1978){Lemaire}, {Skumanich}, {Artzner},
  {Gouttebroze}, {Vial}, {Bonnet}, \& {McWhirter}}]{lemaire1978}
{Lemaire}, P., {Skumanich}, A., {Artzner}, G., {et~al.} 1978, in Bulletin of
  the American Astronomical Society, Vol.~10, 440

\bibitem[{{Lemen} {et~al.}(2012){Lemen}, {Title}, {Akin}, {Boerner}, {Chou},
  {Drake}, {Duncan}, {Edwards}, {Friedlaender}, {Heyman}, {Hurlburt}, {Katz},
  {Kushner}, {Levay}, {Lindgren}, {Mathur}, {McFeaters}, {Mitchell}, {Rehse},
  {Schrijver}, {Springer}, {Stern}, {Tarbell}, {Wuelser}, {Wolfson}, {Yanari},
  {Bookbinder}, {Cheimets}, {Caldwell}, {Deluca}, {Gates}, {Golub}, {Park},
  {Podgorski}, {Bush}, {Scherrer}, {Gummin}, {Smith}, {Auker}, {Jerram},
  {Pool}, {Soufli}, {Windt}, {Beardsley}, {Clapp}, {Lang}, \&
  {Waltham}}]{lemen2012}
{Lemen}, J.~R., {Title}, A.~M., {Akin}, D.~J., {et~al.} 2012, \solphys, 275,
  17, \dodoi{10.1007/s11207-011-9776-8}

\bibitem[{{Lin} \& {Carlsson}(2015)}]{lin2015}
{Lin}, H.-H., \& {Carlsson}, M. 2015, \apj, 813, 34,
  \dodoi{10.1088/0004-637X/813/1/34}

\bibitem[{{Lites}(1980)}]{lites1980}
{Lites}, B.~W. 1980, \solphys, 68, 327, \dodoi{10.1007/BF00156871}

\bibitem[{{L{\'o}pez Fuentes} \& {Klimchuk}(2018)}]{lopez2018}
{L{\'o}pez Fuentes}, M., \& {Klimchuk}, J.~A. 2018, Boletin de la Asociacion
  Argentina de Astronomia La Plata Argentina, 60, 207

\bibitem[{{L{\'o}pez Fuentes} \& {Klimchuk}(2022)}]{lopez2022}
---. 2022, arXiv e-prints, arXiv:2210.01896.
\newblock \doarXiv{2210.01896}

\bibitem[{{Nicolas} {et~al.}(1982){Nicolas}, {Bartoe}, {Brueckner}, \&
  {Kjeldseth-Moe}}]{nicolas1982}
{Nicolas}, K.~R., {Bartoe}, J. D.~F., {Brueckner}, G.~E., \& {Kjeldseth-Moe},
  O. 1982, \solphys, 81, 253, \dodoi{10.1007/BF00151301}

\bibitem[{{Reale}(2014)}]{reale2014}
{Reale}, F. 2014, Living Reviews in Solar Physics, 11, 4,
  \dodoi{10.12942/lrsp-2014-4}

\bibitem[{{Rottman} {et~al.}(1982){Rottman}, {Orrall}, \&
  {Klimchuk}}]{rottman1982}
{Rottman}, G.~J., {Orrall}, F.~Q., \& {Klimchuk}, J.~A. 1982, \apj, 260, 326,
  \dodoi{10.1086/160257}

\bibitem[{{Sandlin} {et~al.}(1986){Sandlin}, {Bartoe}, {Brueckner}, {Tousey},
  \& {Vanhoosier}}]{sandlin1986}
{Sandlin}, G.~D., {Bartoe}, J. D.~F., {Brueckner}, G.~E., {Tousey}, R., \&
  {Vanhoosier}, M.~E. 1986, \apjs, 61, 801, \dodoi{10.1086/191131}

\bibitem[{{Schou} {et~al.}(2012{\natexlab{a}}){Schou}, {Borrero}, {Norton},
  {Tomczyk}, {Elmore}, \& {Card}}]{schou2012a}
{Schou}, J., {Borrero}, J.~M., {Norton}, A.~A., {et~al.} 2012{\natexlab{a}},
  \solphys, 275, 327, \dodoi{10.1007/s11207-010-9639-8}

\bibitem[{{Schou} {et~al.}(2012{\natexlab{b}}){Schou}, {Scherrer}, {Bush},
  {Wachter}, {Couvidat}, {Rabello-Soares}, {Bogart}, {Hoeksema}, {Liu},
  {Duvall}, {Akin}, {Allard}, {Miles}, {Rairden}, {Shine}, {Tarbell}, {Title},
  {Wolfson}, {Elmore}, {Norton}, \& {Tomczyk}}]{schou2012b}
{Schou}, J., {Scherrer}, P.~H., {Bush}, R.~I., {et~al.} 2012{\natexlab{b}},
  \solphys, 275, 229, \dodoi{10.1007/s11207-011-9842-2}

\bibitem[{{Simnett}(1981)}]{simnett1981}
{Simnett}, G.~M. 1981, in International Cosmic Ray Conference, Vol.~12,
  International Cosmic Ray Conference, 205--227

\bibitem[{{Teriaca} {et~al.}(1999){Teriaca}, {Banerjee}, \&
  {Doyle}}]{teriaca1999}
{Teriaca}, L., {Banerjee}, D., \& {Doyle}, J.~G. 1999, \aap, 349, 636

\bibitem[{{Tripathi} {et~al.}(2009){Tripathi}, {Mason}, {Dwivedi}, {del Zanna},
  \& {Young}}]{tripathi2009}
{Tripathi}, D., {Mason}, H.~E., {Dwivedi}, B.~N., {del Zanna}, G., \& {Young},
  P.~R. 2009, \apj, 694, 1256, \dodoi{10.1088/0004-637X/694/2/1256}

\bibitem[{{Tripathi} {et~al.}(2012){Tripathi}, {Mason}, \&
  {Klimchuk}}]{tripathi2012}
{Tripathi}, D., {Mason}, H.~E., \& {Klimchuk}, J.~A. 2012, \apj, 753, 37,
  \dodoi{10.1088/0004-637X/753/1/37}

\bibitem[{{Tripathi} {et~al.}(2020){Tripathi}, {Nived}, {Isobe}, \&
  {Doyle}}]{TriNID_2020}
{Tripathi}, D., {Nived}, V.~N., {Isobe}, H., \& {Doyle}, G.~G. 2020, \apj, 894,
  128, \dodoi{10.3847/1538-4357/ab8558}

\bibitem[{{Wilhelm} {et~al.}(1995){Wilhelm}, {Curdt}, {Marsch}, {Sch{\"u}hle},
  {Lemaire}, {Gabriel}, {Vial}, {Grewing}, {Huber}, {Jordan}, {Poland},
  {Thomas}, {K{\"u}hne}, {Timothy}, {Hassler}, \& {Siegmund}}]{wilhelm1995}
{Wilhelm}, K., {Curdt}, W., {Marsch}, E., {et~al.} 1995, \solphys, 162, 189,
  \dodoi{10.1007/BF00733430}

\bibitem[{{Woodgate} {et~al.}(1980){Woodgate}, {Tandberg-Hanssen}, {Bruner},
  {Beckers}, {Brandt}, {Henze}, {Hyder}, {Kalet}, {Kenny}, {Knox},
  {Michalitsianos}, {Rehse}, {Shine}, \& {Tinsley}}]{woodgate1980}
{Woodgate}, B.~E., {Tandberg-Hanssen}, E.~A., {Bruner}, E.~C., {et~al.} 1980,
  \solphys, 65, 73, \dodoi{10.1007/BF00151385}

\bibitem[{{Young} {et~al.}(2007){Young}, {Del Zanna}, {Mason}, {Dere}, {Landi},
  {Landini}, {Doschek}, {Brown}, {Culhane}, {Harra}, {Watanabe}, \&
  {Hara}}]{young2007}
{Young}, P.~R., {Del Zanna}, G., {Mason}, H.~E., {et~al.} 2007, \pasj, 59,
  S857, \dodoi{10.1093/pasj/59.sp3.S857}

\end{thebibliography}
\end{document}